**High Curie temperatures in ferromagnetic Cr-doped AlN thin films**

D. Kumar,

Electronic Science Department, Kurukshetra University, Kurukshetra 136 119, India

J. Antifakos, M. G. Blamire and Z. H. Barber

Dept. of Materials Science and Metallurgy, University of Cambridge,

Pembroke Street, Cambridge CB2 3QZ, UK

$Al_{1-x}Cr_xN$ thin films with $0.02 \leq x \leq 0.1$ were deposited by reactive co-sputtering onto c-plane (001) sapphire. Room-temperature ferromagnetism with a coercive field of 85 Oe was observed in samples with chromium contents as low as $x = 0.027$ (2.7%). With increasing Cr content the mean magnetic moment is strongly suppressed, with a maximum saturation moment of 0.62 and 0.71 $\mu_B$ per Cr atom at 300 and 50 K respectively. We show that the Curie temperature of $Al_{1-x}Cr_xN$ for $x = 0.027$ is greater than 900 K.

Corresponding author. E-mail: mb52@cam.ac.uk

Since the discovery of carrier-induced ferromagnetism in InMnAs and GaMnAs,[1,2] there has been rapidly growing interest in the properties and potential applications of dilute magnetic semiconductors (DMS). The most significant limitation to application of these materials is the low Curie temperature ($T_C$) generally observed in DMS based on traditional compound semiconductors. This has prompted many groups to investigate alternatives including magnetically-doped oxides such as $ZnO$[3,4] $SnO_2$[5] and $TiO_2$[6].

Theoretical predictions of room temperature ferromagnetism in 3d-transition-metal-doped GaN indicated that Mn and Cr-doped GaN are promising candidate DMS, with Cr-doped GaN likely to exhibit the most stable ferromagnetic states[7]. Dietl *et al.* predicted from a mean field model that Mn-doped GaN should have a $T_C$ higher than room temperature.[8] Based on local-density functional calculations, the exchange interaction between magnetic dopants and III-V compounds may explain the maximum observed critical temperature at different impurity concentrations.[9]

Recently, room temperature ferromagnetism has been experimentally observed in Mn- and Cr- doped GaN and AlN.[10-14] The $T_C$ for (x = 0.03 - 0.05) Mn-doped GaN films has been reported to be over 750 K[15] and ferromagnetic behavior with a $T_C$ higher than 400 K has been reported for Cr-doped GaN.[16] Previous studies on AlN have reported a large variation in the value of mean magnetic moment of Cr-doped AlN films depending on the technique of deposition and doping concentration,[13-14] but it is clear that high Cr doping levels (x ≥ 0.1) result in segregation. To date, samples with varying Cr composition at levels (x < 0.05) have not been studied. In this letter, we report the growth of $Al_{1-x}Cr_xN$ films and show that high $T_C$s can be obtained at low doping levels.

Thin film samples of $Al_{1-x}Cr_xN$, were deposited on c-plane (001) sapphire substrates by reactive D.C. co-sputtering from an Al and a Cr target in a UHV sputtering chamber surrounded by a liquid-nitrogen filled jacket with a base pressure below $3\times10^{-7}$ Pa. A sputtering gas of 70% $N_2$ / 30% Ar at a total pressure of 3 Pa was used. The temperature of the platinum resistive heater, on which the substrates were placed, was held at 940 °C (using an optical pyrometer); the actual substrate temperature was approximately 150 °C below this.

The substrate heater was placed between the Al and Cr targets such that a series of $Al_{1-x}Cr_xN$ films with different Cr compositions could be fabricated on a set of $10 \times 5$ mm$^2$ substrates in one deposition run. In this geometry $Al_{1-x}Cr_xN$ films with Cr (x < 0.027) could not be deposited because of limitations on the relative powers which could be applied to the Al and Cr targets. The concentration of Cr in the films was estimated by energy dispersive X-ray analysis (EDX) of films deposited on to silica substrates placed at either end of the heater in each run; previous experiments had shown that our geometry gives an essentially linear composition spread and so the Cr-concentration of each sample could be accurately calculated from its position on the heater. Film structure and crystallographic texture were characterized using a Philips X-ray diffractometer (XRD) equipped with a Cu source. The magnetic measurements were carried out with the magnetic field parallel to the sample using a Princeton Measurement Corporation vibrating sample magnetometer fitted with a furnace and a cryostat which enabled measurements between 10 K and 800 K.

The thickness of films varied between 400 nm and 600 nm. Atomic Force Microscopy was used to characterize the surface morphology; the mean and root mean square roughness values for a typical film were 4.8 and 6.1 nm respectively. The surface topography suggested columnar growth.

Our growth conditions resulted in a preferred (002) orientation for pure AlN, i.e. c-axis normal to the plane of the film. Fig. 1 shows $2\theta$-$\theta$ XRD scans of $Al_{1-x}Cr_xN$ thin films with different Cr levels deposited at the same temperature on c-plane sapphire. It can be seen that all the films are predominantly (002) oriented, but that with increasing $x$ the intensities of other reflections increase relative to the (002) diffraction peak. For x = 0.027 and 0.035 only the (002) reflection at 36.1° can be observed; the sample with x = 0.046 the first indication of the (101) AlN peak appears; and for x = 0.061 this peak can clearly be observed. The same trend was observed in other deposition runs: preferred c-axis orientation dropped with increasing Cr content. No peaks corresponding to any other possible phase, such as CrN, $Cr_2N$ or $CrO_2$, were observed.

The magnetization versus magnetic field (*M-H*) loops for $Al_{1-x}Cr_xN$ with x = 0.027 and x = 0.061 at 300 K are shown in Fig. 2; both loops are strongly hysteretic, but it can be seen that the total moment for x = 0.061 is substantially lower than for x = 0.027. *M-H* loops measured for x = 0.027 at different temperatures (20 K- 800 K) were used to calculate the saturation magnetization moment ($M_S$), magnetic remanent magnetization ($M_R$) per Cr atom, are shown in Fig. 3[17]. Coercive field values as a function of temperature are also shown in this figure. It can be seen that the film is still strongly ferromagnetic at the highest measurement temperature of our system, 800 K.

As expected, the data show a progressive reduction in $M_S$ with increasing temperature but, as shown in Fig. 3, we have also consistently observed a reduction in $M_S$ values with reducing temperature below than 50K. The effective magnetic moment per Cr atom, deduced from *M-H* loops measured at room temperature, as a function of x is plotted in Fig. 4. The data show a consistent reduction in total moment (inset of Fig. 4) with increasing Cr concentration.

The major difficulty in making AlN ferromagnetic is the low solubility limit[18] for magnetic impurities such as Mn or Cr. However, there appears to be a narrow processing window in terms of deposition temperature and doping concentration at which effective ferromagnetic doping of AlN is possible and which also avoids second phase formation and maintains the preferred growth orientation.[11] Our films have been deposited under good vacuum conditions and so ferromagnetic $CrO_2$ formation, possible for the highly doped films of Yang *et al*.[14] is extremely unlikely. XRD analysis does not detect any other phase, although this does not necessarily preclude the presence of nano-clusters of Cr or Cr-compounds.

The remanent magnetization for the 2.7% doped sample was found to decrease with increasing temperature (>50 K); however the films were strongly hysteric at all temperatures. The coercive field and effective magnetization also decreased with increasing temperature. The relatively low signal to noise ratio at high temperatures prevented the use of Arrott plots to determine $T_C$; however, the *M-H* measurements made at higher temperatures clearly indicate that $T_C$ is higher than 800K. Extrapolation of the linear dependence of the remanent magnetization to higher temperatures implies a $T_C$ well above 900 K, and probably closer to 1000 K.

The rapid decrease in effective saturation magnetic moment $M_S$ per Cr atom with increasing Cr content may be associated with the decline in crystal quality with increasing Cr content or may be due to enhanced antiferromagnetic coupling at higher Cr concentrations. Alternatively, it is possible that this decrease is a consequence of an increasing chemical instability with higher Cr concentration which may lead to clustering. This decrease in total moment with increasing doping concentration is similar to that already reported for Mn-doped GaN films[19]. The reason for the decrease in $M_S$ (< 50 K) is not understood, but may also indicate a competing antiferromagnetic coupling.

At present, there seems to be no single theoretical model which can explain the ferromagnetism in all DMS materials. Ferromagnetism based on free-carrier mediated models predicted for Mn-doped III-V semiconductors may not be applicable to AlN as it is highly resistive at room temperature. It has recently been proposed that small Cr-N clusters could be strongly ferromagnetic,[20] and so could account for the magnetic moment and high $T_C$s in Cr-doped III-V nitrides. However, this appears to be inconsistent with our data, in which ferromagnetism is strongest at the lowest Cr concentrations and, as noted by Hori *et al.*[15], the small size of the clusters necessary for ferromagnetism seems certain to imply a superparamagnetic behaviour which is inconsistent with the hysteresis observed even at the highest temperatures in our films.

In conclusion, we have deposited ferromagnetic Cr-doped AlN films with strong (002) texture on c-plane sapphire, showing high saturation magnetization moments of 0.62 and 0.71$\mu_B$/Cr atom at 300 and 50 K respectively, for Cr contents as low as 2.7%. The magnetization measurements carried out at higher temperatures indicate that the Curie temperature of Cr-doped AlN is higher than 900 K. The most immediate application for high temperature ferromagnetic AlN may be in spin electronic devices as a spin-filtering magnetic tunnel barrier.[11]

DK thanks ACU, London for awarding a Commonwealth Fellowship to support this research. The authors thank Dr Madhav Adyam and Dr. Dennis Leung for their assistance with VSM measurements and for useful discussions.

Figure captions

FIG. 1 X-ray diffraction patterns of $Al_{1-x}Cr_xN$ samples with varying Cr composition ($0.02 \leq x \leq 0.1$), deposited on c-plane (001) sapphire at heater temperature of 940 °C.

FIG. 2 Magnetic field dependence of the magnetization of $Al_{1-x}Cr_xN$ films for x = 0.027 and 0.061 at 300 K.

FIG. 3 Temperature dependence of the effective saturation magnetic moment ($M_S$), magnetic remanent magnetization ($M_R$), and coercive field ($H_c$) of $Al_{1-x}Cr_xN$ with x = 0.027.

FIG. 4 Effective magnetic moment per Cr atom for $Al_{1-x}Cr_xN$ film as a function of Cr concentration measured at 300K; different growth runs are distinguished by different symbols. The inset shows the same data expressed as moment per unit volume.

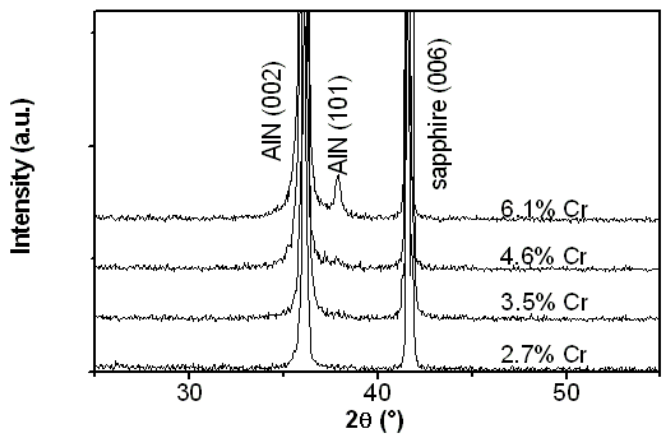

FIG. 1. – Kumar *et al*

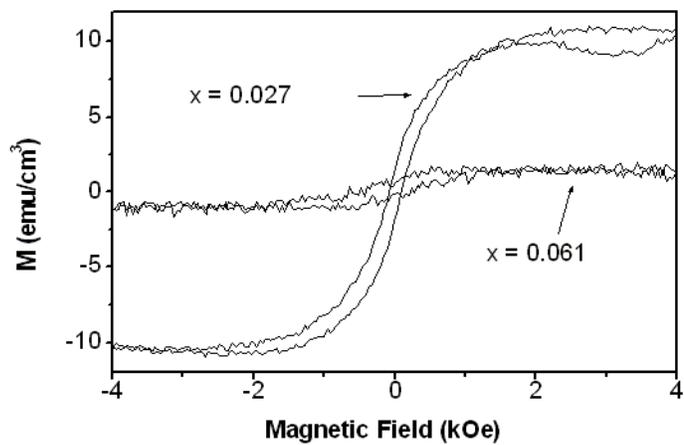

FIG. 2. – Kumar *et al*

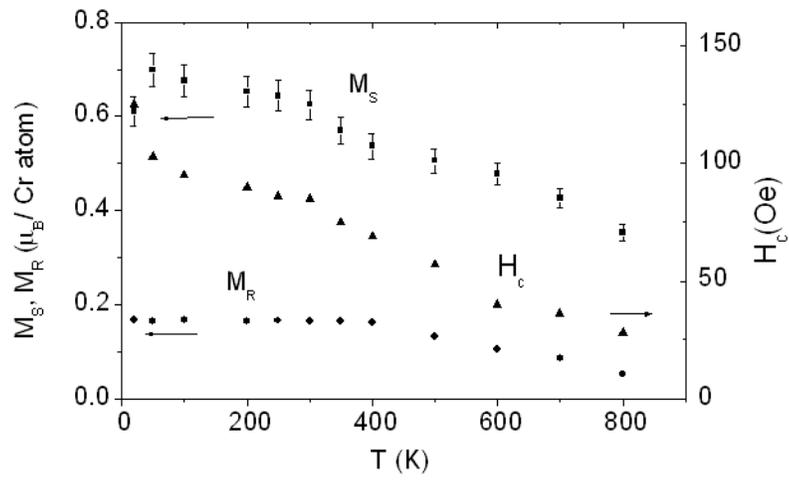

FIG. 3. – Kumar *et al*

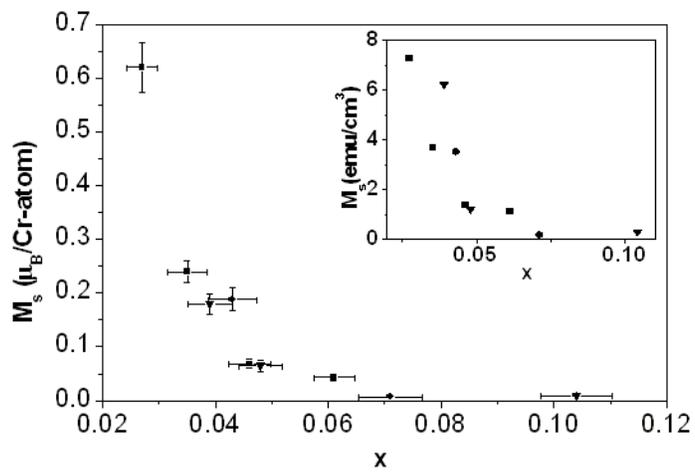

FIG. 4. – Kumar *et al*